\documentstyle[twoside]{article}
\input{psfig.sty}

\catcode`\@=11
\long\def\@makefntext#1{
\protect\noindent \hbox to 3.2pt {\hskip-.9pt  
$^{{\eightrm\@thefnmark}}$\hfil}#1\hfill}               

\def\thefootnote{\fnsymbol{footnote}}
\def\@makefnmark{\hbox to 0pt{$^{\@thefnmark}$\hss}}    
        
\def\ps@myheadings{\let\@mkboth\@gobbletwo
\def\@oddhead{\hbox{}
\rightmark\hfil\eightrm\thepage}   
\def\@oddfoot{}\def\@evenhead{\eightrm\thepage\hfil
\leftmark\hbox{}}\def\@evenfoot{}
\def\sectionmark##1{}\def\subsectionmark##1{}}

\oddsidemargin=\evensidemargin
\renewcommand{\thefootnote}{\fnsymbol{footnote}}

\newcounter{sectionc}\newcounter{subsectionc}\newcounter{subsubsectionc}
\renewcommand{\section}[1] {\vspace{12pt}\addtocounter{sectionc}{1} 
\setcounter{subsectionc}{0}\setcounter{subsubsectionc}{0}\noindent 
        {\tenbf\thesectionc. #1}\par\vspace{5pt}}
\renewcommand{\subsection}[1] {\vspace{12pt}\addtocounter{subsectionc}{1} 
        \setcounter{subsubsectionc}{0}\noindent 
        {\bf\thesectionc.\thesubsectionc. {\kern1pt \bfit #1}}\par\vspace{5pt}}
\renewcommand{\subsubsection}[1] {\vspace{12pt}\addtocounter{subsubsectionc}{1}
        \noindent{\tenrm\thesectionc.\thesubsectionc.\thesubsubsectionc.
        {\kern1pt \tenit #1}}\par\vspace{5pt}}
\newcommand{\nonumsection}[1] {\vspace{12pt}\noindent{\tenbf #1}
        \par\vspace{5pt}}
\newcommand{\gsim}{\mathrel{\rlap{\raisebox{.3ex}{$>$}}
    \raisebox{-.6ex}{$\sim$}}}
\newcommand{\lsim}{\mathrel{\rlap{\raisebox{.3ex}{$<$}}
    \raisebox{-.6ex}{$\sim$}}}

\newcounter{appendixc}
\newcounter{subappendixc}[appendixc]
\newcounter{subsubappendixc}[subappendixc]
\renewcommand{\thesubappendixc}{\Alph{appendixc}.\arabic{subappendixc}}
\renewcommand{\thesubsubappendixc}
        {\Alph{appendixc}.\arabic{subappendixc}.\arabic{subsubappendixc}}

\renewcommand{\appendix}[1] {\vspace{12pt}
        \refstepcounter{appendixc}
        \setcounter{figure}{0}
        \setcounter{table}{0}
        \setcounter{lemma}{0}
        \setcounter{theorem}{0}
        \setcounter{corollary}{0}
        \setcounter{definition}{0}
        \setcounter{equation}{0}
        \renewcommand{\thefigure}{\Alph{appendixc}.\arabic{figure}}
        \renewcommand{\thetable}{\Alph{appendixc}.\arabic{table}}
        \renewcommand{\theappendixc}{\Alph{appendixc}}
        \renewcommand{\thelemma}{\Alph{appendixc}.\arabic{lemma}}
        \renewcommand{\thetheorem}{\Alph{appendixc}.\arabic{theorem}}
        \renewcommand{\thedefinition}{\Alph{appendixc}.\arabic{definition}}
        \renewcommand{\thecorollary}{\Alph{appendixc}.\arabic{corollary}}
        \renewcommand{\theequation}{\Alph{appendixc}.\arabic{equation}}
        \noindent{\tenbf Appendix \theappendixc #1}\par\vspace{5pt}}
\newcommand{\subappendix}[1] {\vspace{12pt}
        \refstepcounter{subappendixc}
        \noindent{\bf Appendix \thesubappendixc. {\kern1pt \bfit #1}}
        \par\vspace{5pt}}
\newcommand{\subsubappendix}[1] {\vspace{12pt}
        \refstepcounter{subsubappendixc}
        \noindent{\rm Appendix \thesubsubappendixc. {\kern1pt \tenit #1}}
        \par\vspace{5pt}}

\newcommand{\textlineskip}{\baselineskip=13pt}
\newcommand{\smalllineskip}{\baselineskip=10pt}

\def\eightcirc{
\begin{picture}(0,0)
\put(4.4,1.8){\circle{6.5}}
\end{picture}}
\def\eightcopyright{\eightcirc\kern2.7pt\hbox{\eightrm c}} 

\newcommand{\copyrightheading}[1]
        {\vspace*{-2.5cm}\smalllineskip{\flushleft
        {\footnotesize International Journal of Modern Physics A, #1}\\
        {\footnotesize $\eightcopyright$\, World Scientific Publishing
         Company}\\
         }}

\def\abstracts#1#2#3{{
        \centering{\begin{minipage}{4.5in}\baselineskip=10pt\footnotesize
        \parindent=0pt #1\par 
        \parindent=15pt #2\par
        \parindent=15pt #3
        \end{minipage}}\par}} 

\renewenvironment{thebibliography}[1]
        {\frenchspacing
         \ninerm\baselineskip=11pt
         \begin{list}{\arabic{enumi}.}
        {\usecounter{enumi}\setlength{\parsep}{0pt}
         \setlength{\leftmargin 12.7pt}{\rightmargin 0pt} 
         \setlength{\itemsep}{0pt} \settowidth
        {\labelwidth}{#1.}\sloppy}}{\end{list}}

\newcounter{itemlistc}
\newcounter{romanlistc}
\newcounter{alphlistc}
\newcounter{arabiclistc}

\newcommand{\fcaption}[1]{
        \refstepcounter{figure}
        \setbox\@tempboxa = \hbox{\footnotesize Fig.~\thefigure. #1}
        \ifdim \wd\@tempboxa > 5in
           {\begin{center}
        \parbox{5in}{\footnotesize\smalllineskip Fig.~\thefigure. #1}
            \end{center}}
        \else
             {\begin{center}
             {\footnotesize Fig.~\thefigure. #1}
              \end{center}}
        \fi}

\newcommand{\tcaption}[1]{
        \refstepcounter{table}
        \setbox\@tempboxa = \hbox{\footnotesize Table~\thetable. #1}
        \ifdim \wd\@tempboxa > 5in
           {\begin{center}
        \parbox{5in}{\footnotesize\smalllineskip Table~\thetable. #1}
            \end{center}}
        \else
             {\begin{center}
             {\footnotesize Table~\thetable. #1}
              \end{center}}
        \fi}

\def\@citex[#1]#2{\if@filesw\immediate\write\@auxout
        {\string\citation{#2}}\fi
\def\@citea{}\@cite{\@for\@citeb:=#2\do
        {\@citea\def\@citea{,}\@ifundefined
        {b@\@citeb}{{\bf ?}\@warning
        {Citation `\@citeb' on page \thepage \space undefined}}
        {\csname b@\@citeb\endcsname}}}{#1}}

\newif\if@cghi
\def\cite{\@cghitrue\@ifnextchar [{\@tempswatrue
        \@citex}{\@tempswafalse\@citex[]}}
\def\citelow{\@cghifalse\@ifnextchar [{\@tempswatrue
        \@citex}{\@tempswafalse\@citex[]}}
\def\@cite#1#2{{$\null^{#1}$\if@tempswa\typeout
        {IJCGA warning: optional citation argument 
        ignored: `#2'} \fi}}

\def\pmb#1{\setbox0=\hbox{#1}
        \kern-.025em\copy0\kern-\wd0
        \kern.05em\copy0\kern-\wd0
        \kern-.025em\raise.0433em\box0}

\def\fnt#1#2{\footnotetext{\kern-.3em
        {$^{\mbox{\scriptsize #1}}$}{#2}}}

\def\fpage#1{\begingroup
\voffset=.3in
\thispagestyle{empty}\begin{table}[b]\centerline{\footnotesize #1}
        \end{table}\endgroup}

\def\runninghead#1#2{\pagestyle{myheadings}
\markboth{{\protect\footnotesize\it{\quad #1}}\hfill}
{\hfill{\protect\footnotesize\it{#2\quad}}}}
\font\tenrm=cmr10
\font\tenit=cmti10 
\font\tenbf=cmbx10
\font\bfit=cmbxti10 at 10pt
\font\ninerm=cmr9

\font\eightrm=cmr8

\def\qed{\hbox{${\vcenter{\vbox{                        
   \hrule height 0.4pt\hbox{\vrule width 0.4pt height 6pt
   \kern5pt\vrule width 0.4pt}\hrule height 0.4pt}}}$}}

\renewcommand{\thefootnote}{\fnsymbol{footnote}}        

\def\ket#1{|#1\rangle}

\begin{document}

\runninghead{Neutrino Physics: An experimental overview $\ldots$}
{Neutrino Physics: An experimental overview $\ldots$}

\normalsize\textlineskip
\thispagestyle{empty}
\setcounter{page}{1}

\copyrightheading{}                     

\vspace*{0.88truein}

\fpage{1}
\centerline{\bf NEUTRINO PHYSICS:}
\vspace*{0.035truein}
\centerline{\bf AN EXPERIMENTAL OVERVIEW}
\vspace*{0.37truein}
\centerline{\footnotesize KATE SCHOLBERG\footnote{Present address: Department of Physics, Massachusetts Institute of Technology, Cambridge, MA
02139.}}
\vspace*{0.015truein}
\centerline{\footnotesize\it Department of Physics, Boston University}
\baselineskip=10pt
\centerline{\footnotesize\it Boston, MA 02215,
USA}

\vspace*{0.21truein}
\abstracts{The field of neutrino physics is currently very
exciting, with several recent results pointing to new physics.
I will give an overview of the current experimental situation, focusing
primarily on neutrino oscillation results.  The
data are not entirely consistent however, and puzzles remain.
I will then review the new experiments which are poised to solve the
outstanding puzzles.}{}{}

\vspace*{1pt}\textlineskip      
\section{Introduction}    
\vspace*{-0.5pt}
\noindent

Neutrinos are now known to have mass and to mix.  It is still not
known exactly how they fit into the theoretical picture, however. In
the baseline Standard Model, neutrinos are massless, and extensions to
the Standard Model are required to explain their small masses and the
observed oscillation phenomena.  These observations can be
accommodated in GUTs, and perhaps more exotic scenarios such as extra
dimensions$^{\ref{albright}}$).  Neutrinos are also important for
cosmology, and play a role in the early universe. Even a tiny $\nu$
mass is sufficient for neutrinos to make up a significant portion of
the total mass of the Universe.  To further understanding, we must
turn to experiment: we must quantify knowledge of masses, map the
mixing matrix, and determine whether CP violating phases are present
for neutrino as well as quark mixing We must determine whether sterile
neutrinos exist. Another important question is whether neutrinos are
Majorana or Dirac (i.e.  whether or not they are equivalent to their
antiparticles), which can be probed experimentally by searching for
double beta decay, or dipole moments of neutrinos.

Here I will focus on one significant aspect of experimental neutrino
physics: the search for neutrino oscillations.  But before proceeding,
I will pause to welcome the last neutrino to the family: $\tau$
leptons from the interactions of $\nu_{\tau}$'s have finally been
observed by the DONUT (Direct Observation of the Nu Tau)
experiment at Fermilab$^{\ref{donut}}$.  This experiment used a prompt
$\nu_{\tau}$ beam and an active emulsion detector to look for the
``kink'' indicating a $\nu_{\tau}$-induced $\tau$ lepton decaying
$\sim$~millimeters from the interaction vertex.  DONUT collaborators
find four ``long decay'' events (which have a measured $\tau$ parent
to the decay), where they expect $4.1 \pm 1.4$ signal events, and
$0.41 \pm 0.15$ background events. More analyses are in progress.

\pagebreak

\textheight=7.8truein
\setcounter{footnote}{0}
\renewcommand{\thefootnote}{\alph{footnote}}

\section{Neutrino Mass and Oscillations}
\noindent

\subsection{Neutrino Oscillations}

Neutrino oscillations arise from straightforward quantum 
mechanics.  We assume that the $N$ neutrino flavor states 
$\ket{\nu_f}$, which participate
in the weak interactions, are superpositions of the mass
states $\ket{\nu_i}$, and are related by a unitary mixing matrix:
\begin{equation}
\ket{\nu_f} = \sum_{i=1}^{N} U_{fi} \ket{\nu_i}.
\end{equation}
For the two-flavor case, assuming
relativistic neutrinos, it can easily be shown that
the probability for flavor transition is given by
\begin{equation}\label{eq:oscprob}
P(\nu_f\rightarrow\nu_g)=\sin^22\theta\sin^2(1.27\Delta m^2 L/E),
\end{equation}
for $\Delta m^2 \equiv m_2^2-m_1^2$ in eV$^2$.
$L$ (in km) is the distance traveled by the neutrino and $E$ in GeV
is its energy.
Several comments are in order:

\begin{itemize}
\item Note that in this equation
the parameters that experimenters try to measure
(and theorists try to derive)
are $\sin^22\theta$ and $\Delta m^2$.
$L$ and $E$ depend on the experimental situation.  
\item The
neutrino oscillation probability depends on mass squared
differences, not absolute masses.

\item The neutrino mixing matrix can be 
easily generalized to three flavors, and the transition probabilities
computed in a straightforward way.

\item For three flavors,
there are only two \textbf{independent} $\Delta m^2$ values.  

\item If the mass states are not nearly degenerate,
one is often in a ``decoupled'' regime where it is possible to
describe the oscillation as effectively two-flavor, i.e. following
an equation similar to~\ref{eq:oscprob}, with effective mixing angles and
mass squared differences.  I will assume a two-flavor description of the
mixing in most cases here.

\item ``Sterile'' neutrinos, $\nu_s$, with no normal
weak interactions, are possible in many theoretical scenarios (for
instance, as an isosinglet state in a GUT$^{\ref{albright}}$).
  
\end{itemize}

\section{The Three Experimental ``Hints''}
\noindent

There are currently three experimental ``hints'' of neutrino
oscillations. These indications are summarized in Table 1.  I will now
examine the current status of each of these observations.

\begin{table}[htbp]
\tcaption{Experimental evidence for neutrino oscillations.}
\centerline{\footnotesize\smalllineskip
\begin{tabular}{c c c c c c}\\
\hline
$\nu$ source & Experiments & Flavors & $E$ & $L$  & $\Delta m^2$ sensitivity (eV$^2$)\\
\hline
Sun & Chlorine & $\nu_e \rightarrow \nu_x$ & 5-15 MeV & 10$^{8}$ km &  $10^{-12}-10^{-10}$\\
    & Gallium  &                           &          &           & or $10^{-6}-10^{-4}$ \\
    & Water Cherenkov &                    &          &           & \\
\hline
Cosmic ray & Water Cherenkov & $\nu_{\mu} \rightarrow \nu_x$ & 0.1-100 GeV & $10-10^5$ km & $10^{-2}-10^{-3}$ \\
showers   & Iron calorimeter & & & & \\
          & Upward muons & & & & \\
\hline
Accelerator & LSND & $\bar{\nu}_{\mu} \rightarrow \bar{\nu}_e$   & 15-50 MeV & 30 m & 0.1-1 \\\ 
           &       &  $\nu_{\mu} \rightarrow \nu_e$ &  20-200 MeV &\ & \\ 
\hline\\ 
\end{tabular}}
\end{table}

\subsection{Solar Neutrinos}
\noindent

The deficit of solar neutrinos was the
first of these ``hints'' to be observed.  The solar neutrino energy
spectrum is well predicted, and depends primarily on weak physics,
being rather insensitive to any solar physics.  Three types
of solar neutrino detectors (chlorine, gallium and water Cherenkov),
with sensitivity at three different
energy thresholds, together observe an energy-dependent suppression
which cannot be explained by any solar model 
(standard or non-standard)$^{\ref{bahcall}}$.

\begin{figure}[h]
\begin{center}
\leavevmode
\psfig{figure=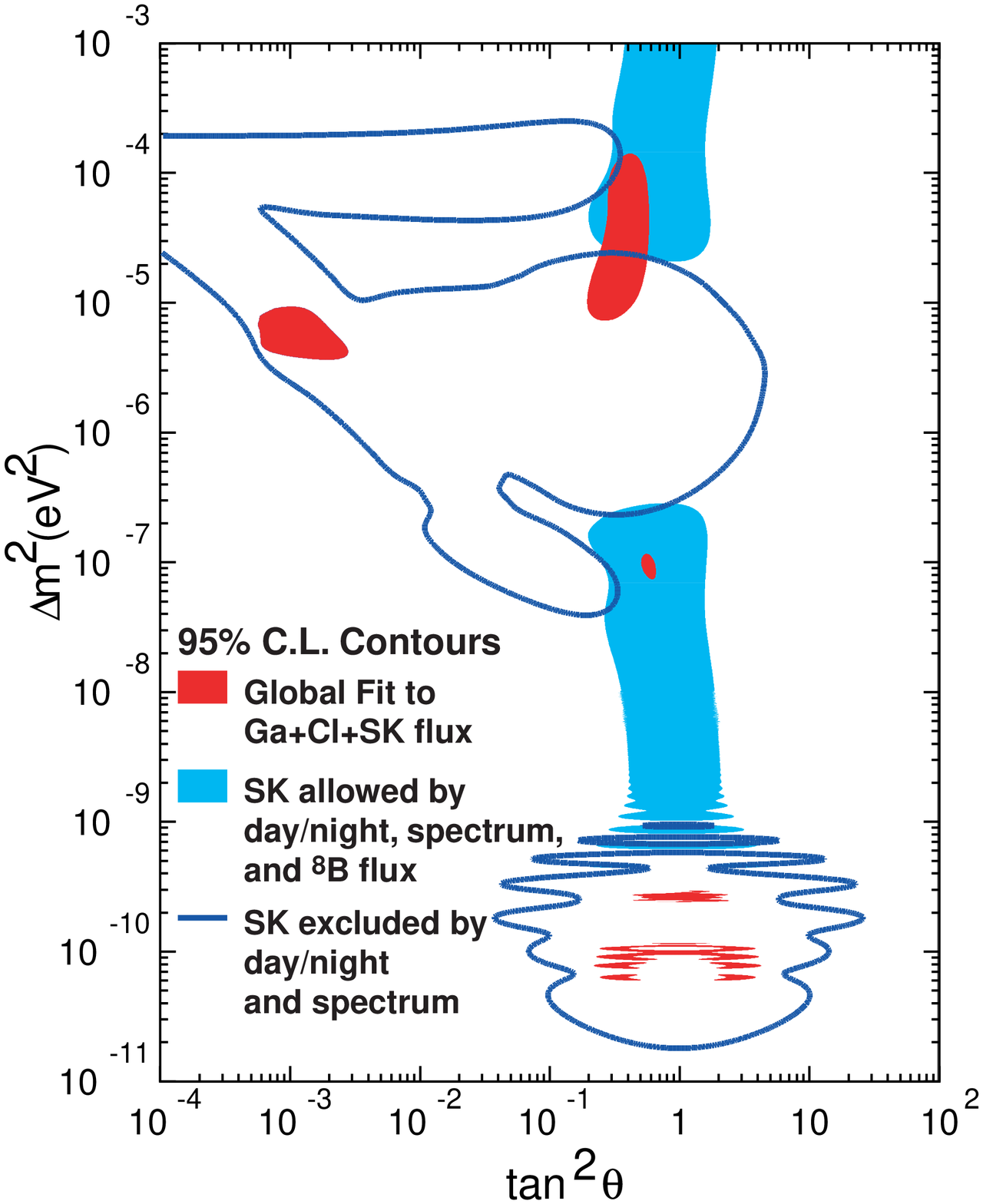,height=4in}
\end{center}
\fcaption{Solar neutrino parameter space: the dark
areas show the global flux fit solutions.  The interiors
of the dark lines indicate Super-K's excluded
regions; the light shaded areas indicate Super-K's
allowed regions.}\label{fig:solar_param}
\end{figure}

The observed suppression in all three experiments can be explained by
neutrino oscillation at certain values of $\Delta m^2$ and mixing
angle: see Figure~\ref{fig:solar_param}.  The allowed regions at
higher values of $\Delta m^2$ (``small mixing angle'', ``large mixing
angle'' and ``low'') are those for which matter effects in the Sun
come into play. The solutions at
very small $\Delta m^2$ values are the vacuum
oscillation solutions.  Figure~\ref{fig:solar_param} shows the mixing
angle axis plotted as $\tan^2\theta$, rather than as the more
conventional $\sin^22\theta$, to make evident the difference between
$0<\theta<\pi/4$ and $\pi/4<\theta<\pi/2$: these regions are not
equivalent when one considers matter effects$^{\ref{murayama}}$.

The measured energy spectrum of solar neutrinos is a potential
``smoking gun'' for neutrino oscillations: a distortion from the
expected shape would be hard to explain by other than non-standard weak
physics.  The latest Super-K solar neutrino spectrum shows no evidence
for distortion$^{\ref{sksolar}}$.  Another ``smoking gun'' solar neutrino
measurement is the day/night asymmetry: electron neutrinos may be
regenerated in the Earth from their oscillated state for certain
oscillation parameters.  The latest measured Super-K day/night
asymmetry is $\frac{D-N}{(D+N)/2}=-0.034 \pm 0.022
^{+0.013}_{-0.012}$: regeneration is therefore a relatively small
effect, if it is present at all.  Together, the energy
spectrum and day/night observations place
strong constraints on solar neutrino parameters.  In particular,
Figure~\ref{fig:solar_param} shows the Super-K results overlaid on the
global flux fit parameters: large mixing angles are favored, and the
small mixing angle and vacuum solutions from the global flux fit are
disfavored at 95\% C.L..  In addition, all global flux fit $\nu_e
\rightarrow \nu_s$ solutions are disfavored at 95\% C.L..  See also
reference~\ref{krastev} from these proceedings.


\subsection{Atmospheric Neutrinos}
\noindent

\begin{figure}[htbp]
\begin{center}
\leavevmode
\psfig{figure=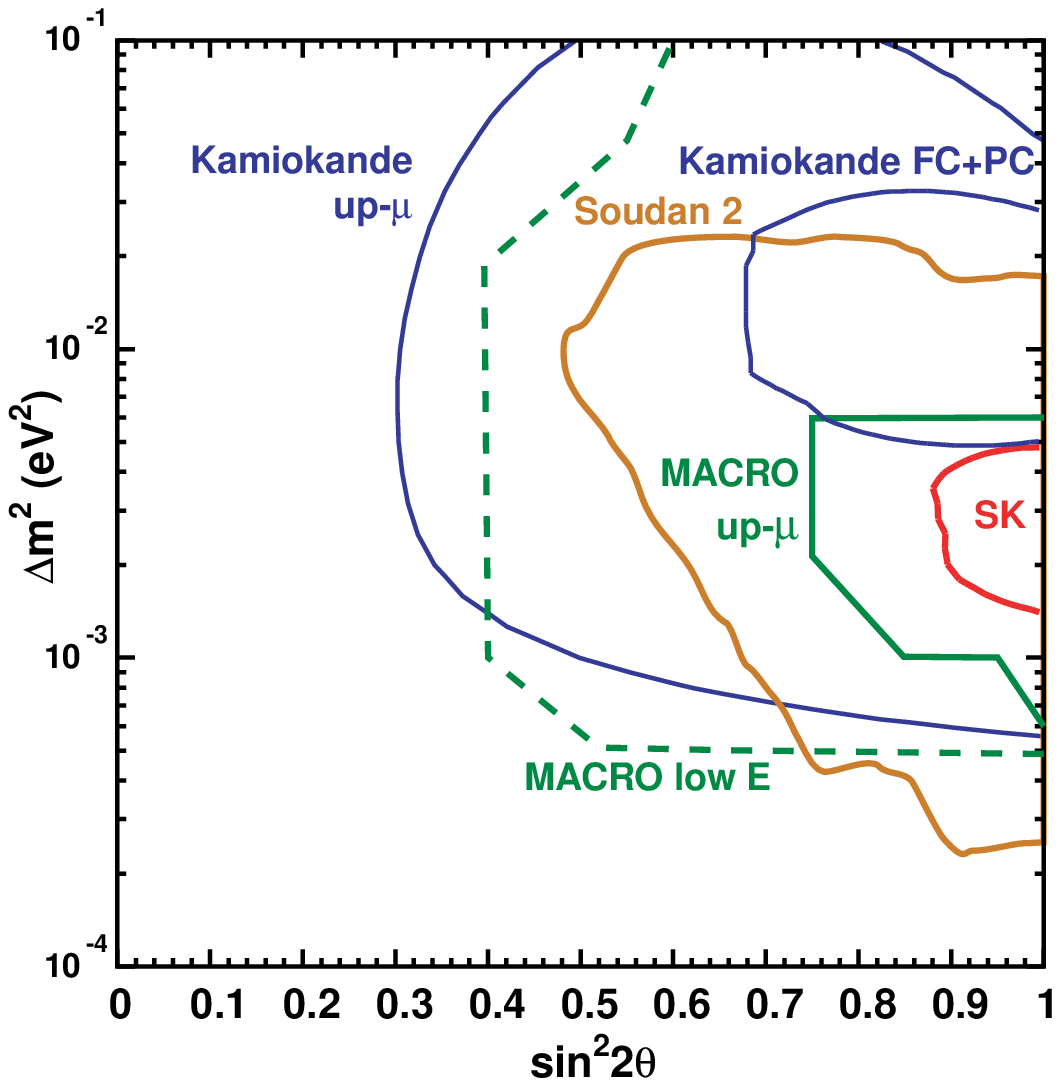,height=2.9in}
\end{center}
\fcaption{Atmospheric neutrino oscillation parameter space.
Parameters to the right of the contours shown are allowed by
the experiments.}\label{fig:atmnu_space}
\end{figure}

Atmospheric neutrinos are produced by collisions of cosmic rays
with the upper atmosphere.  Energies
are in the range from 0.1~GeV to 100~GeV.  Atmospheric
neutrinos can be observed coming from all directions.
At neutrino energies $\gsim$ 1~GeV, 
for which the geomagnetic field has very little effect on
the primaries,
by geometry the neutrino flux should be up-down symmetric.
Although the absolute flux prediction has $\sim$15\% uncertainty,
the flavor ratio (about two muon neutrinos for
every electron neutrino) is known quite robustly, since it
depends on the well-understood decay chain 
$\pi \rightarrow \mu \nu_{\mu} \rightarrow e \nu_e \bar{\nu}_\mu$.
The Super-K result of 1998$^{\ref{nuosc}}$ showed a highly significant
deficit of $\nu_\mu$ events from below, with an energy
and pathlength dependence as expected from equation~\ref{eq:oscprob}.  
The most recent data constrain the two-flavor
$\nu_{\mu}\rightarrow \nu_{\tau}$ oscillation parameters to a region
as shown in Figure~\ref{fig:atmnu_space}.  The latest results from
Soudan 2$^{\ref{soudan}}$ (an iron tracker) and from
MACRO's$^{\ref{macro}}$ upward-going muon sample are consistent with
the Super-K data.

Recently, Super-K has also been able to shed some light on which
flavors are involved in the muon neutrino disappearance.  Assuming a
two-flavor oscillation, the missing $\nu_{\mu}$'s could have
oscillated into either $\nu_e$, $\nu_{\tau}$ or $\nu_s$.  The
oscillation cannot be pure $\nu_{\mu} \rightarrow \nu_{e}$, because
there is no significant excess of $\nu_e$ from below.  In addition,
the CHOOZ$^{\ref{chooz}}$ and Palo Verde$^{\ref{paloverde}}$ experiments
have ruled out disappearance of reactor $\bar{\nu}_e$, for similar
parameters. Three-flavor fits of the Super-K data have also been done;
small mixing to $\nu_e$ is allowed$^{\ref{sk3flav}}$.

It difficult to test the $\nu_{\mu}\rightarrow\nu_{\tau}$ hypothesis
directly. Super-K expects only tens of charged current (CC)
$\nu_{\tau}$ interactions in the current sample, and the products of
such interactions in the detector are nearly indistinguishable from
other atmospheric neutrino events.  However, recently Super-K has
employed two strategies to distinguish
$\nu_{\mu}\rightarrow\nu_{\tau}$ from
$\nu_{\mu}\rightarrow\nu_{s}$$^{\ref{tauvss}}$.  First, one can look
for an angular distortion of high-energy neutrinos due to matter
effects of sterile neutrinos propagating in the Earth: unlike
$\nu_{\tau}$'s, sterile neutrinos do not exchange $Z^0$'s with matter
in the Earth, resulting in an MSW-like effect that effectively
suppresses oscillation.  The effect is more pronounced at higher
energies.  Such distortion of the high-energy event angular distribution 
is not observed.  Second, one can
look at neutral current (NC) events in the detector: if oscillation is
to a sterile neutrino, the neutrinos ``really disappear'' and do not
interact via NC.  A NC-enriched sample of multiple-ring Super-K events
shows no deficit of up-going NC events.  Together, these measurements
exclude two-flavor $\nu_{\mu}\rightarrow\nu_{s}$ at 99\% C. L., for
all parameters allowed by the Super-K fully-contained events.

\subsection{LSND}

\begin{figure}[htbp]
\begin{center}
\leavevmode
\psfig{figure=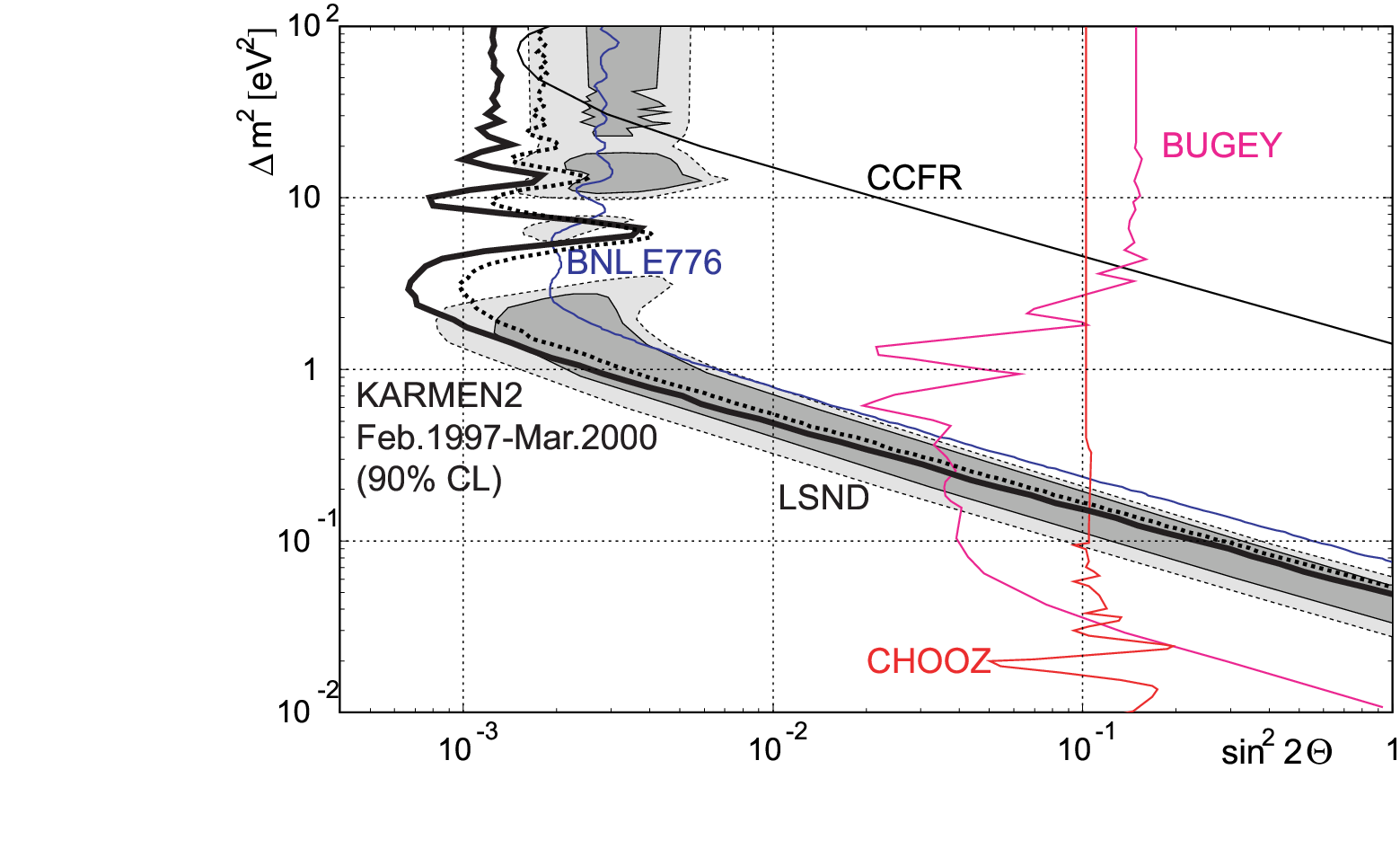,bbllx=2pt,bblly=183pt,bburx=607pt,bbury=598pt,height=2.3in}
\end{center}
\fcaption{LSND parameter space.  The shaded region indicates the
parameters allowed by LSND. The region to the right of the dark
line is excluded by KARMEN.  The Bugey reactor experiment 
excludes the large mixing angle parameters. Plot from
K. Eitel$^{\ref{karmen}}$.}\label{fig:lsnd_space}
\end{figure}

The third oscillation hint is the only ``appearance'' observation: the
Liquid Scintillator Neutrino Detector (LSND) experiment at Los Alamos
has observed an excess of $\bar{\nu}_e$ events$^{\ref{lsnd}}$ from a
beam which should contain only $\bar{\nu}_{\mu}$, $\nu_e$ and
$\nu_{\mu}$ from positive pion and muon decay at rest.  The result is
interpreted as $\sim$20-50~MeV $\bar{\nu}_{\mu}$'s oscillating over a
30~m baseline.  The LSND collaboration also observes an excess of
higher energy (60-200~MeV) electrons$^{\ref{lsnd2}}$, presumably due to
$\nu_e$'s oscillated from $\nu_{\mu}$'s from pion decay in flight.
This observation is
consistent with the oscillation hypothesis.  See
Figure~\ref{fig:lsnd_space} for the corresponding allowed region in
parameter space.  The large mixing angle part of this range is ruled out
by reactor experiments.

An experiment at Rutherford-Appleton Laboratories called KARMEN, which
has roughly similar neutrino oscillation sensitivity (17.5~m baseline)
as does LSND, does not however confirm the LSND result$^{\ref{karmen}}$.
This detector expects fewer signal events than does LSND, but has a
stronger background rejection due to the pulsed nature of their
neutrino beam.  However, due to somewhat different sensitivity,
KARMEN's lack of observation of $\bar{\nu}_e$ appearance cannot rule out all
of the parameter space indicated by LSND: see
Figure~\ref{fig:lsnd_space}.

\subsection{Accelerator Searches at High $\Delta m^2$}

Another region of parameter space which has been recently been
explored is in the regime where $\Delta m^2 > 1$ eV$^2$.  This is an
interesting region for cosmological reasons: neutrino masses in 
this range could have a significant effect on cosmological
models.  However, no signal has been found in this region.  In
particular, the CHORUS and NOMAD experiments at CERN, which 
observed the same $\nu_{\mu}$ beam 
($\langle{E}_{\nu}\rangle\sim$ 25 GeV, baseline
$\sim$ 600~m) with quite different detector technologies (emulsion and
electronic tracking, respectively), have both excluded
$\nu_{\mu}\rightarrow\nu_{\tau}$ at 90\% C.L.  for 
$\Delta m^2 \gsim 0.7$ eV$^2$ at maximal mixing,
down to $\sin^22\theta \gsim 5-8\times 10^{-4}$ for
$\Delta m^2 \gsim 100$ eV$^2$ $^{\ref{chorus},\ref{nomad}}$.


\section{Where Do We Stand?}

Now we can step back and view the big picture.  Where do we stand?
The current experimental picture for the three oscillation signal
indications can be summarized:

\begin{itemize}

\item For solar neutrino parameter space ($\nu_e \rightarrow \nu_x$):
Super-K's day/night and energy spectrum data disfavor small mixing
angle and vacuum solutions; large mixing is favored.  Pure
$\nu_{\mu}\rightarrow \nu_s$ is disfavored.

\item For atmospheric neutrino parameter space, evidence
from  Super-K, Soudan 2 and MACRO 
is very strong for $\nu_{\mu} \rightarrow \nu_x$.  Furthermore,
Super-K's data favor the $\nu_{\mu} \rightarrow \nu_{\tau}$
hypothesis over the $\nu_{\mu} \rightarrow \nu_s$ one.

\item The LSND indication of $\nu_{\mu} \rightarrow \nu_x$ still
stands; KARMEN does not rule out all of LSND's allowed parameters.
\end{itemize}
\begin{figure}[htbp]
\begin{center}
\leavevmode
\psfig{figure=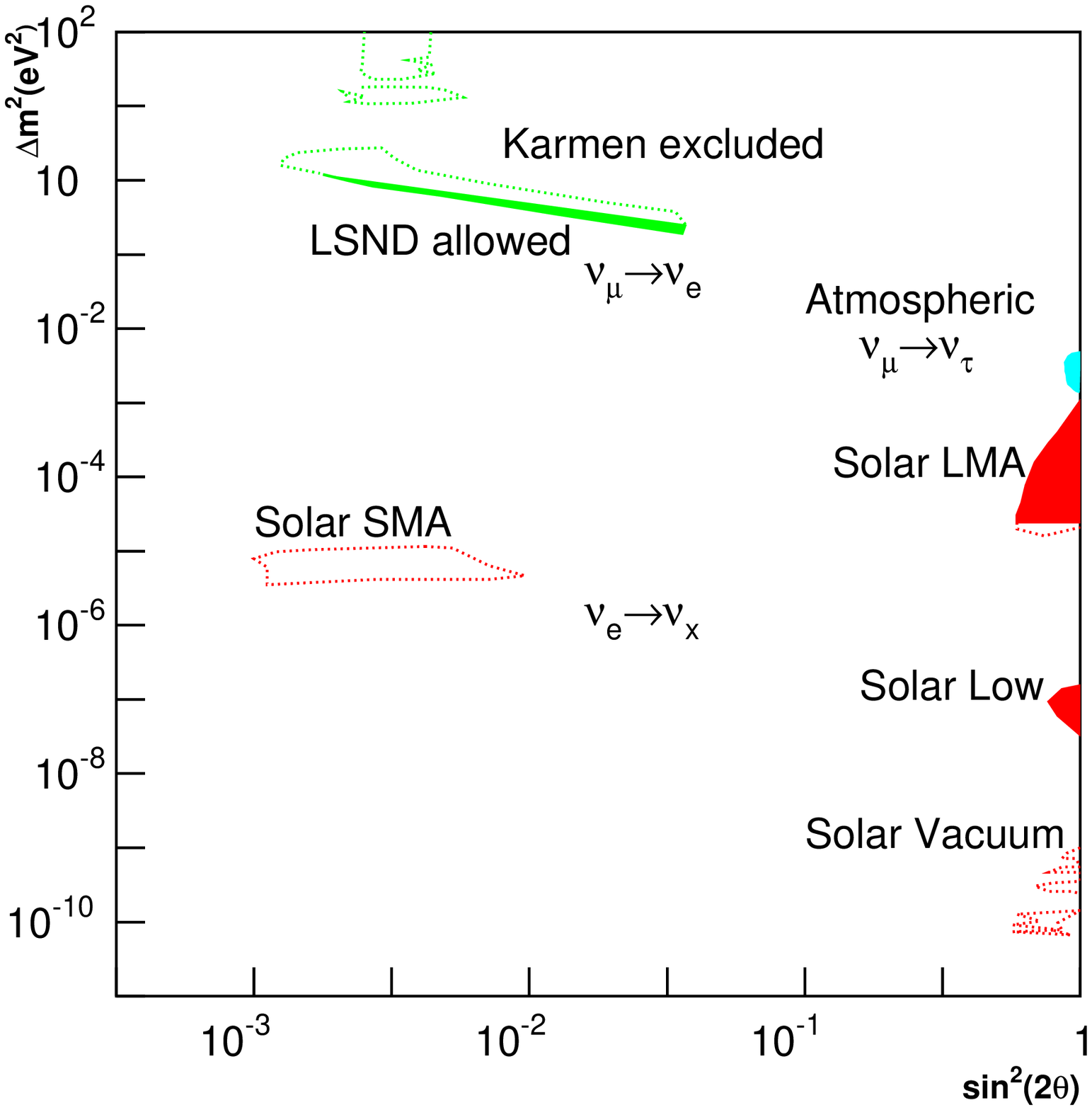,height=3.2in,width=2.2in}
\end{center}
\fcaption{Overall parameter space: the current picture.
Solid regions are allowed by the latest data; dotted regions encompass
regions allowed until recently.}
\end{figure}

What do these data mean?  There is an obvious problem. Under the
assumption of three generations of massive neutrinos, there are only
two independent values of $\Delta m^2$: we must have $\Delta
m_{13}^2=\Delta m_{12}^2 + \Delta m_{23}^2$.  However, we have three
measurements which give $\Delta m^2$ values of three different orders
of magnitude.  So, if each hint represents two-flavor mixing, then
something must be wrong.  Three-flavor fits are
unsatisfactory~$^{\ref{4flavfit}}$.  One way to wriggle out of this
difficulty is to introduce another degree of freedom in the form of a
sterile neutrino (or neutrinos). 
(We cannot introduce another light active neutrino,
due to the Z$^0$ width measurements from LEP, which constrain the
number of active neutrinos to be three$^{\ref{lep}}$: any new light
neutrino must be sterile).  Although pure mixing into $\nu_s$ is now
disfavored by Super-K solar and atmospheric neutrino results, a
sterile neutrino is still viable as part of some four-flavor
mixing$^{\ref{4flavfit}}$.  Of course, it is also possible that some
of the data are wrong or misinterpreted.  Clearly, we need more
experiments to clarify the situation.

\section{What's Next?}

So what's next?  I will cover the next experiments
for each of the interesting regions of parameter space.

\subsection{LSND Neutrino Parameter Space}

The next experiment to investigate the LSND parameter space will
be BooNE (Booster Neutrino Experiment). This will look at
$\sim$ 1 GeV neutrinos from the 8~GeV booster at Fermilab, at
a baseline of about 500~m (with a second experiment planned
at longer baseline if an oscillation signal is seen).
This experiment can test $\nu_{\mu}$ disappearance, but is
primarily designed to test $\nu_{\mu}\rightarrow \nu_e$ at about
the same $L/E$ as LSND. Since the energy is higher, 
and the backgrounds are different, systematics
will presumably be different from those at LSND.  BooNE,
which will start in December 2001, expects
about 500 oscillation signal events per year at LSND parameters,
and expects to cover all of LSND space in one year of 
running$^{\ref{boone}}$.

\subsection{Solar Neutrino Parameter Space}

A great variety of experiments are working towards an understanding of
solar neutrinos.  The first new experiment among these is the SNO
(Sudbury Neutrino Observatory) experiment$^{\ref{sno}}$.  This
experiment contains 1~kton of D$_2$O, aiming to measure CC and NC
breakup reactions, as well as elastic scattering (ES). Information
about solar neutrino oscillations will be obtained by observing the
distortion of the CC energy spectrum, as well as CC/NC and CC/ES
ratios.  SNO has been taking data for about one year, and this summer
SNO reported its first CC data$^{\ref{snodata}}$.  The first NC data
and oscillation analyses are expected soon.

Two new low energy threshold scintillator detectors are on the very
near horizon.  Borexino at the Gran Sasso Laboratory in
Italy$^{\ref{borexino}}$ is a 300~ton scintillator experiment, with
very low radioactive background.  It will start in 2002.  Borexino's
aim is to see the solar 0.86~MeV $^7$Be line.  Measurement of
day/night and seasonal variations of this line should give very good
sensitivity to vacuum and ``low'' solar neutrino oscillation
solutions.  KamLAND is a 1~kton scintillator detector at the Kamioka
mine in Japan$^{\ref{kamland}}$.  This detector, which should start
operation in 2001, intends to study the large mixing angle part of
solar parameter space with
\textit{reactor} neutrinos: it will measure the sum of the fluxes of
$\bar{\nu}_e$ from reactors within about 500~km in Japan and Korea.
KamLAND also hopes to study solar neutrinos directly.  
There are also a number of innovative
new solar neutrino experiments
aiming to look at the very low energy pp solar flux$^{\ref{lownu}}$.

\subsection{Atmospheric Neutrino Parameter Space}

For atmospheric neutrino parameter space, the next steps are the
``long-baseline'' experiments, which aim to test the atmospheric
neutrino oscillation hypothesis directly with an artificial beam of
neutrinos.  

The first of these is the K2K (KEK to Kamioka) experiment$^{\ref{k2k}}$,
which started in March 1999, and which saw the first artificial
long-distance
neutrinos in June 1999.  K2K sends a beam of $\langle{E}_{\nu}\rangle\sim$1~GeV
$\nu_{\mu}$ 250~km across Japan to the Super-K experiment. 
K2K can look for $\nu_{\mu}$ disappearance (the beam
energy is not high enough to make significant $\tau$'s).  
Preliminary results do show a deficit of observed neutrinos:
40.3$^{+4.7}_{-4.6}$ beam events in the fiducial volume are expected, but
only 27 are seen, which disfavors no oscillations at the 2$\sigma$
level.  About 25\% of K2K data has been taken.

The MINOS experiment$^{\ref{minos}}$, to start in 2003, will send a
beam from Fermilab to Soudan, with a beam energy of $3-20$ GeV, and a
baseline of 730~km.  The far detector is a magnetic iron tracker.
CNGS (Cern Neutrinos to Gran Sasso)$^{\ref{cngs}}$, a $\sim$20~GeV
$\nu_{\mu}$ beam from CERN to the Gran Sasso 730~km away, will start
in 2005.  These long baseline experiments, thanks to high beam
energies and high statistics, can expand the search beyond $\nu_{\mu}$
disappearance: they can also look for $\nu_e$ appearance at small
mixing, and possibly also for $\tau$ appearance.

A possibility on the farther horizon for atmospheric neutrino
parameter space is a neutrino factory: a muon storage ring would
produce copious, and well-understood, 20-50~GeV neutrinos from muon
decay$^{\ref{nufact}}$.  Detectors at 3000-7000~km
baselines could study atmospheric neutrino oscillation parameters with
unprecedented precision, could possibly measure $\nu_{e}-\nu_{\tau}$
mixing, and could even perhaps find CP violation (although this would
have to be disentangled from matter effects in the Earth).  Other
possibilities being explored are more conventional $\nu$ beams of high
intensity$^{\ref{lepviol}}$.



\begin{figure}[htbp]
\begin{center}
\leavevmode
\psfig{figure=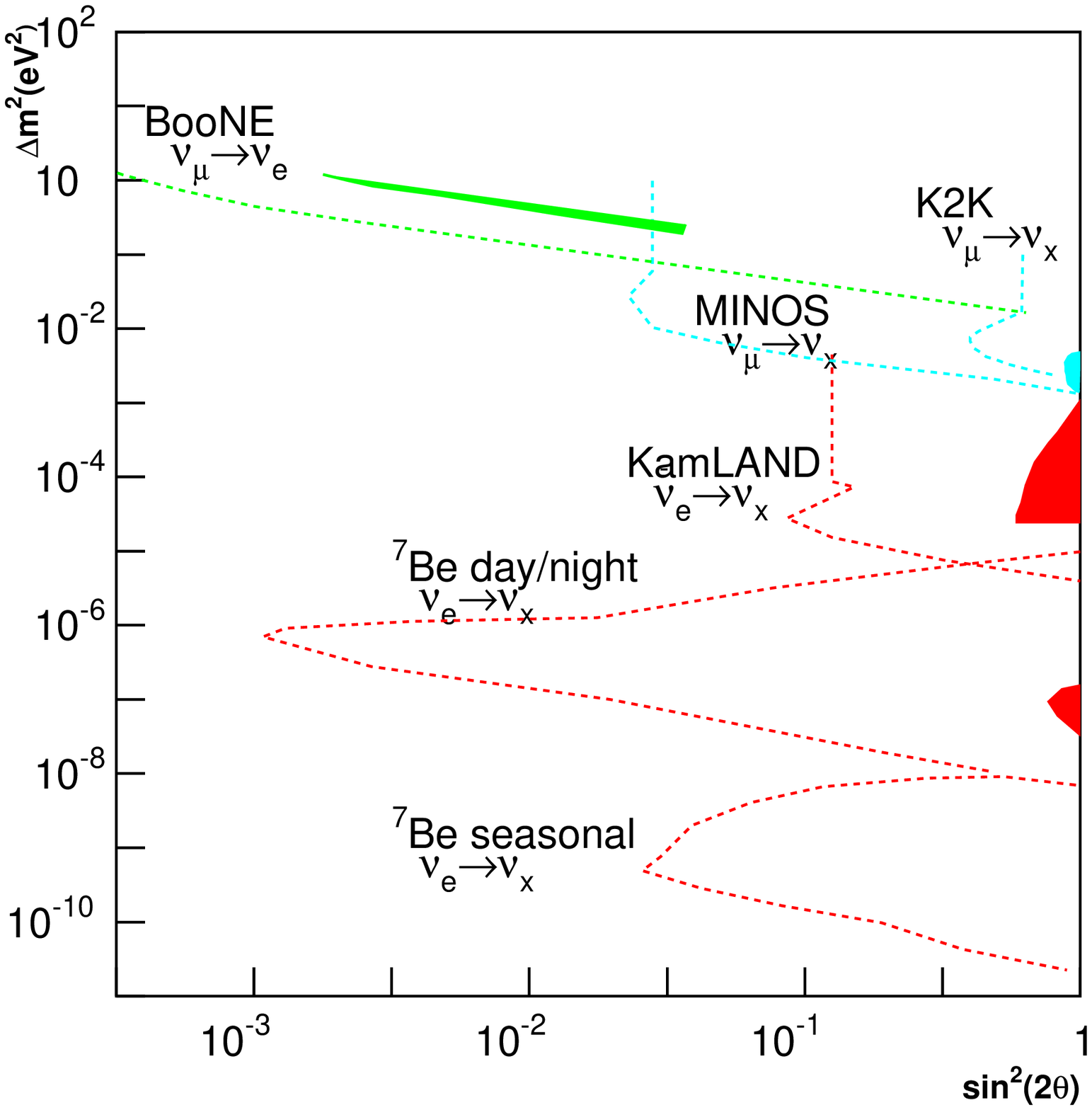,height=3.3in,width=2.2in}
\end{center}
\fcaption{Sensitivities of the next neutrino oscillation experiments.}\label{whatsnext}\label{fig:whatsnext}
\end{figure}

Figure~\ref{fig:whatsnext} summarizes the reach of the next
generation of neutrino experiments.

\section{Other Neutrino Experiments}

This review has focused on neutrino oscillation studies.  However, I
do not wish to leave the impression that oscillations are all of
neutrino physics.  There are many other ways of clarifying the
neutrino picture, and I will mention some of them briefly here.

\begin{itemize}

\item \textit{Kinematic Absolute Mass Limits:} As noted above,
neutrino oscillation measurements say nothing about absolute masses of
the mass states.  Although the kinematic mass searches are difficult,
the concept is simple: look for missing energy.  The traditional
tritium beta decay spectrum endpoint experiments now have limits for
absolute $\nu_e$ mass of $\lsim$2.5~eV/$c^2$, and prospects are good
for improvement down to
$\lsim$0.5~eV/$c^2$~$^{\ref{mainz},\ref{troitsk}}$.  The $\nu_{\mu}$
and $\nu_{\tau}$ mass limits are currently
190~keV/$c^2$~$^{\ref{numukin}}$ and 15.5~MeV/$c^2$~$^{\ref{nutaukin}}$
respectively.

\item \textit{Double Beta Decay:} Another way of getting at absolute
neutrino mass is to look for neutrinoless double beta decay, $(N,Z)
\rightarrow (N-2,Z+2)+e^-+e^-$.  Such a decay is only possible if the
neutrino has mass, and is Majorana.  The current lowest mass limits
from non-observation of double beta decay are about $\langle
m_{\nu}\rangle_e<0.2$ eV/$c^2$~$^{\ref{hm},\ref{ejiri}}$.  Many new
experiments are being built, and prospects are good for improving the
limits down to 0.01~eV/$c^2$~$^{\ref{newbb}}$.

\item \textit{Neutrino Magnetic Moment:}
A non-zero non-transition magnetic moment implies that neutrinos are
Dirac and not Majorana.  The best limits on neutrino magnetic
moment$^{\ref{magmom}}$ are astrophysical.  The current best
laboratory limits are $\mu_{\nu}\lsim10^{-10}\mu_B$.  The MUNU
experiment, which measures elastic scattering of reactor
$\bar{\nu}_e$, will improve this limit by an order of
magnitude$^{\ref{munu}}$.

\item \textit{Supernova Neutrinos:}
A supernova is a ``source of opportunity'' for neutrino physics: we
can expect a giant burst of neutrinos of all flavors from a core
collapse in our Galaxy about once every 30~years.  Many of the large
neutrino experiments -- Super-K, SNO, Borexino, KamLAND, LVD and
AMANDA -- are sensitive$^{\ref{snnu}}$.  We may be able to extract
absolute mass information from time of flight measurements, as well as
some oscillation information by looking for spectral distortion.

\item \textit{High Energy Neutrino Astronomy:}
Long string water Cherenkov detectors (AMANDA, Baikal, Antares,
Nestor) in ice and water, are embarking on a new era of high energy
neutrino astronomy$^{\ref{longstring}}$.  From the $\gsim$10~GeV neutrino
signals in these detectors, we can learn about astrophysical sources
of neutrinos, as well as about oscillation at high energy.

\item \textit{Cosmology:} Finally, neutrinos play a significant role
in cosmology$^{\ref{cosmology}}$.  They may make up some
non-negligible component of the dark matter (although this is now
thought to be small, in order to be consistent with galactic
structure formation).  Cosmological considerations constrain the sum of
absolute masses of the states to $\sim$10 eV or less.  The ultra
low energy (1.95~K) big-bang relic neutrinos$^{\ref{relic}}$,
which are expected to permeate the universe with a number density of
113 cm$^{-3}$, remain an experimental challenge.

\end{itemize}

\section{Summary}

In summary, I will list recent highlights of 
experimental neutrino physics:

\begin{itemize}

\item The DONUT experiment has directly detected $\nu_{\tau}$-induced
$\tau$'s for the first time.
\item Super-K's solar neutrino results favor large mixing
and disfavor the small mixing and vacuum oscillation solutions of the global
flux fit.  Pure $\nu_e \rightarrow \nu_s$ mixing is also disfavored
for all parameters.
\item The SNO experiment has presented its first CC data.  
\item Super-K atmospheric neutrinos favor $\nu_{\mu} \rightarrow
\nu_{\tau}$ over $\nu_{\mu} \rightarrow \nu_s$ based on angular distributions of high energy events,
and of a NC-enriched sample.
\item With 25\% of data taken,
K2K's deficit of beam $\nu_{\mu}$'s disfavors the no-oscillation hypothesis.
\item The LSND $\nu_{\mu} \rightarrow \nu_e$ ($\bar{\nu}_{\mu} \rightarrow \bar{\nu}_e$)
signal is not
ruled out by the latest KARMEN data, and will be tested by BooNE.

\end{itemize}

It is now quite certain that neutrinos have mass and mix.  The details
of the picture are getting more clear, but many puzzles remain.  The
data taken together are inconsistent with the mixing of three families
of active neutrinos, and we still do not know whether all results are
correct, or whether wider assumptions are needed.  We do not have
enough information to know how to fit the observations into extensions
of the Standard Model.  However, many new experiments are poised to
answer these questions, and clearly, the next ten years will be
extremely interesting.


\nonumsection{References}

\end{document}